# On the Nature of the Bonding in Metal-Silane σ-Complexes

*G. Sean McGrady,*[*,†] *Peter Sirsch,*[†] *Nicholas P. Chatterton,*[§] *Andreas Ostermann,*[¶] *Carlo Gatti,*[≠] *Sandra Altmannshofer,*[‡] *Verena Herz,*[‡] *Georg Eickerling,*[‡] *and Wolfgang Scherer,*[*,‡]

Department of Chemistry, University of New Brunswick, 30 Dineen Drive, Fredericton, N.B. E3B 6E2, Canada, Department of Health and Human Sciences, London Metropolitan University, 166-220 Holloway Road, London N7 8DB, U.K., Forschungsneutronenquelle Heinz Maier-Leibnitz (FRM-II), Technische Universität München, D-85747 Garching, Germany, CNR-ISTM, Istituto di Scienze e Tecnologie Molecolari, via C. Golgi 19, 20133 Milano, Italy, and Lehrstuhl für Chemische Physik und Materialwissenschaften, Universität Augsburg, Universitätsstr. 1, D-86159 Augsburg, Germany

* To whom correspondence should be addressed. E-mail: smcgrady@unb.ca (G.S.M.); wolfgang.scherer@physik.uni-augsburg.de (W.S.).

[†] University of New Brunswick.

[‡] Universität Augsburg.

[§] London Metropolitan University.

[≠] CNR-ISTM.

[¶] FRM-II, Technische Universität München.

The nature of metal silane $\sigma$-bond interaction has been investigated in several key systems by a range of experimental and computational techniques. The structure of [Cp'Mn(CO)$_2$($\eta^2$-HSiHPh$_2$)] **1** has been determined by single crystal neutron diffraction, and the geometry at the Si atom is shown to approximate to a trigonal bipyramid; salient bond distances and angles are Mn–H(1) 1.575(14), Si–H(1) 1.806(14), Si–H(2) 1.501(13) Å and H(1)–Si–(H2) 148.5(8)°. This complex is similar to [Cp'Mn(CO)$_2$($\eta^2$-HSiFPh$_2$)] **2**, whose structure and bonding characteristics have recently been determined by charge density studies based on high-resolution X-ray and neutron diffraction data. The geometry at the Si atom in these $\sigma$-bond complexes is compared with that in other systems containing hypercoordinate silicon. The Mn–H distances for **1** and **2** in solution have been estimated using NMR $T_1$ relaxation measurements, giving a value of 1.56(3) Å in each case, in excellent agreement with the distances deduced from neutron diffraction. DFT calculations have been employed to explore the bonding in the Mn-H-Si unit in **1** and **2** and in the related system [Cp'Mn(CO)$_2$($\eta^2$-HSiCl$_3$)] **3**. These studies support the idea that the oxidative addition of a silane ligand to a transition metal center may be described as an asymmetric process in which the Mn–H bond is formed at an early stage, while both the establishment of the Mn–Si bond and also the activation of the $\eta^2$-coordinated Si-H moiety are controlled by the extent of Mn $\rightarrow$ $\sigma$*(*X*–Si–H) back-donation, which increases with increasing electron-withdrawing character of the *X* substituent *trans* to the metal-coordinated Si-H bond. This delocalized molecular orbital (MO) approach is complemented and supported by combined experimental and theoretical charge density studies: the source function $S(\mathbf{r},\Omega)$, which provides a measure of the relative importance of each atom's contribution to the density at a specific reference point **r** clearly shows that all three atoms of the Mn($\eta^2$-SiH) moiety contribute to a very similar extent to the density at the Mn-Si bond critical point, in pleasing agreement with the MO model. Hence, we advance a consistent and unifying concept which accounts for the degree of Si-H activation in these silane $\sigma$-bond complexes.

# 1. Introduction

The chemistry of σ-bond complexes formed by $\eta^2$-coordination of a ligand E–H bond (E = C, Si, H, B, Sn or Ge) to a transition metal (TM) center has been the subject of intense interest over the past three decades.[1,2,3,4] These systems provide an insight into the activation of E–H bonds by TM centers;[1] a series of complexes may be viewed as "snapshots" at various stages along the reaction coordinate for oxidative addition of the E–H bond to the metal. Silane σ-bond complexes were the first to be isolated and recognized as such in 1969.[5] They currently represent the second largest class of σ-bond complexes behind molecular hydrogen systems, and are of additional importance as a model for their more ephemeral alkane σ-bond cousins and for C–H activation.[6] In the 1980s, Schubert and co-workers prepared and studied a wide range of silane complexes derived from the 16-electron TM fragment [($\eta^5$-$C_5R_5$)Mn(CO)L] (L = $PR_3$ or CO).[7] In this body of work, the σ-bond was characterized primarily on the basis of the distance between the coordinated hydrogen and silicon atoms (*ca.* 1.8 Å), and from the value of the NMR coupling constant (20 < *J* < 60 Hz) between these two atoms. Silane σ-bond complexes have formed the subject of several detailed reviews over the past decade.[8,9,10]

Most of the structural evidence gleaned to date on silane σ-bond complexes has come from X-ray diffraction studies, the one exception being the complex [Cp'Mn(CO)$_2$($\eta^2$-HSiFPh$_2$)] **2** (Cp' = $\eta^5$-$C_5H_4Me$), which was characterized in a neutron diffraction study.[11] The existence of but a single neutron diffraction structure in the literature makes it difficult to draw meaningful conclusions about the Mn($\eta^2$-SiH) bonding interaction, as this is characterized by the parameters *r*(Si–H), *r*(Mn–H), *r*(Mn–Si) and the angle H–Si–*X* (where *X* is the atom *trans* to H), and X-ray diffraction fails to locate the H atom with sufficient accuracy to analyze the Mn–Si–H geometry to any meaningful extent. The geometrical changes at Si that accompany coordination to the TM center are also substantial – from approximately tetrahedral to a distorted trigonal bipyramid (TBP). With its proclivity toward hypercoordination, it is important to understand the geometry at the silicon center in these complexes as well as that at the metal. Furthermore, a number of silane σ-bond complexes exhibit significant secondary interactions between the Si–H moiety and other atoms bound to the TM center. These have been described by Sabo-Etienne *et al.* as *Secondary Interactions between Silicon and Hydrogen Atoms* (SISHA),[12] and by Nikonov as *Inter-ligand Hypervalent Interactions* (IHI).[13]

Prior to this study, no accurate structural data existed for an uncoordinated Si–H bond in a σ-bond complex. In the light of these issues, we have carried out a single crystal neutron diffraction study of [Cp'Mn(CO)$_2$(η$^2$-HSiHPh$_2$)] **1**, the results of which are reported here.

==============

**Scheme 1** near here

==============

As $T_1$ measurements by NMR spectroscopy provide rapid and reliable information on the structure of hydride complexes in the solution phase,[14] we have also chosen to investigate how faithfully the Mn–H–Si geometries of **1** and **2** are retained in solution. This provides an important link: the vast majority of chemical reactions occur in solution, but most structural data pertain to the solid state, where effects such as crystal packing forces can exert a significant influence on the geometry adopted. Solution-phase information is particularly important in the case of TM hydrides, as several members of this class of compounds are known to undergo structural changes in the transition from solid to solution.[15] In this paper, we report the characterization in solution of the Mn–H distance in **1** and **2** by $T_1$ measurements, and we compare these values with their counterparts obtained from single-crystal neutron diffraction studies.

Finally, DFT calculations have been carried out in an attempt to gain a deeper insight into the Mn(η$^2$-SiH) bonding in **1**, **2** and the related system [Cp'Mn(CO)$_2$(η$^2$-HSiCl$_3$)] **3**. Based on photoelectron (PE) spectroscopy studies, complex **3** was previously considered to lie much closer to the oxidative addition product than **1** and **2**,[16,17] and we were particularly interested whether this is also reflected in the electronic structure of these systems. Accordingly, we have examined the molecular orbital (MO) makeup of **1** and **2**, and compared these with those recently published for the cyclopentadienyl analogue of **3**, [CpMn(CO)$_2$($\eta^2$-HSiCl$_3$)] **3a**.[18] A preliminary account of our topological analysis of the electron density in **1-3** was recently reported:[19] here we describe this analysis in more detail, and show that the charge density in each of these systems reveals a Mn–Si–H moiety with the Si–H bond still largely intact.

## 2. Experimental Section

**2.1 Synthesis and Characterization.** All manipulations were carried out under rigorously inert atmosphere conditions using standard Schlenk and glove box techniques.[20] Samples of **1** and **2** were prepared by literature methods.[21] Crystals of **1** suitable for study by neutron diffraction were obtained by dissolving a pure crystalline sample (1.1 g) of the complex in 50 mL of pentane in a Schlenk tube. The vessel was stored at room temperature for 4 d, after which time high quality yellow crystals had formed. Their purity was confirmed by $^1$H NMR spectroscopy. The crystals were collected by filtration and dried under a stream of Ar.

**2.2 Neutron Diffraction Study of 1.** A suitable single crystal of dimensions 3.0 x 2.3 x 2.0 mm was mounted in a sealed quartz capillary and protected from light by a thin Al foil cover. Diffraction data were collected at room temperature using the monochromatic diffractometer BIX-3[22] at the JRR-3M research reactor of the Japanese Atomic Energy Research Institute, Tokai-mura, Japan. The one-circle diffractometer BIX-3 is equipped with a cylindrical neutron imaging plate.[23] An elastically bent, perfect Si(311) single crystal was employed as a monochromator, providing a wavelength of 1.23 Å.[24] Data were collected in three $\omega$-scans (oscillation method, $\Delta\omega = 2.0°$) with 252 frames in total. During the second and third scans, a detachable arc assembly was used. The measurement time per frame was controlled by the monitor counts in front of the crystal, and was approximately 30 min. An initial orientation matrix was determined from 10 frames of the first scan set and refined along with diffractometer constants during integration.[25] The final unit cell parameters were obtained by full-matrix least-squares refinement of 2649 reflections. Integration and scaling of each scan set with the program SCALEPACK[25] resulted in data sets corrected for the effects of crystal decay and absorption. After merging symmetry equivalent and multiply measured reflections with the program SORTAV[26] a unique data set remained which was used for a full-matrix least-squares refinement by minimizing $\Sigma(F_o^2 - F_c^2)^2$ with a SHELXL-97 weighting scheme.[27] The initial atomic coordinates for the heavy atoms were taken from a previously determined X-ray structure and the neutron scattering lengths were taken to be $b_c(C) = 6.646$, $b_c(H) = -3.739$, $b_c(Mn) = -3.73$, $b_c(O) = 5.803$, $b_c(Si) = 4.1419$ fm.[28] During the refinement, difference Fourier maps clearly revealed all of the hydrogen atom positions and all atoms of the asymmetric unit were refined anisotropically. Crystal data and further details of the data collection and the refinement are summarized in Table 1. All geometry calculations were performed with the program PLATON,[29] drawings were generated using ORTEP-3.[30]

=============

**Table 1** near here

=============

**2.3. NMR Spectroscopic Studies.** All NMR experiments were carried out using 5 mm NMR tubes closed with a teflon valve (J. Young, London). These were connected to a Schlenk line via a conical joint of Pyrex glass. The NMR solvent (toluene-$d_8$) was dried and deoxygenated using conventional procedures. $T_1$ relaxation measurements on complexes **1** and **2** were carried out using a Bruker AV400 spectrometer. The conventional inversion-recovery method (180-$\tau$-90)[31] was used to determine $T_1$. $T_{1s}$ relaxation times were measured by applying a selective 180° pulse to the hydride resonance only; the decoupler provided this pulse using a pulse sequence described elsewhere.[32] Calculation of relaxation times was carried out using the non-linear three-parameter fitting routine of the spectrometer software.

In each experiment, the waiting time between each pulse was at least five times the expected relaxation time in order to ensure complete relaxation of the sample, and 16 variable delays were employed. The pulses were calibrated at each temperature, the sample being allowed to equilibrate for at least 10 min before measurements were made. $T_1$ and $T_{1s}$ were measured for complexes **1** and **2** over a range of temperatures covering $T_{1min}$.

*Relaxation Theory.* The detailed theory used to evaluate structural parameters from NMR data has been described elsewhere.[14] We present here only the important equations and the chemical properties that permit the correct application of $T_1$ data to complexes **1** and **2**. It is found that metal-hydrogen dipole-dipole interactions (MHDDI) dominate (along with dipole-dipole relaxation by proximal protons) in classical manganese hydrides.[33] This is because Mn has a large nuclear spin ($I = 5/2$). The contribution of MHDDI to $T_1$ relaxation is defined as described in Eq. 1:

$$1/T_1(\text{Mn}\cdots\text{H}) = (2/15)\, r(\text{Mn–H})^{-6}\, \gamma^2_\text{H}\, \gamma^2_\text{Mn}\, \hbar^2\, I(I+1)\, \{3\tau_c/(1+\omega_\text{H}^2\tau_c^2)$$
$$+ 6\tau_c/[1+(\omega_\text{H}+\omega_\text{Mn})^2\tau_c^2] + \tau_c/[1+(\omega_\text{H}-\omega_\text{Mn})^2\tau_c^2]\} \qquad \textbf{Eq. 1}$$

where $\gamma$, $\omega$, $\hbar$, $\tau_c$ and $I$ have their usual meanings in this context.[31] When $T_1$ reaches a minimum ($T_{1min}$) with respect to temperature, the Mn$\cdots$H internuclear distance can be calculated simply through Eq. 2 ($\nu$ is the $^1$H NMR resonance frequency in MHz):

$$r(\text{Mn–H})\,(\text{Å}) = 2.287[200 T_{1min}(\text{Mn–H})/\nu]^{1/6} \qquad \textbf{Eq. 2}$$

The value of $r$(Mn–H) obtained through Eq. 2 is strictly valid only for isolated Mn and H atoms, and ignores any contribution from proton-proton dipole relaxation or other relaxation processes available to a hydride ligand in a molecular environment. It has been found that Mn–H distances in solution can be reliably determined through Eq. 3 by measuring selective ($T_{1s}$), non-selective ($T_1$) and $T_{1min}$ relaxation times of hydride ligands, when $\omega_H^2 \tau_c^2 \ll 1$.

$$r(\text{Mn–H}) (\text{Å}) = 4.31[(1.4k + 4.47)T_{1min}/\nu]^{1/6} \qquad \text{Eq. 3}$$

Here $k = (f-1)/(\frac{1}{2} - f/3)$, and $f = T_{1s}/T_1$. Eq. 3, unlike Eq. 2, incorporates the effects of other relaxation mechanisms along with MHDDI. A comparison of the results obtained using both Eq. 2 and Eq. 3 gives an indication as to whether MHDDI effects dominate in the relaxation of hydride signal.

For calculation of Mn–H distances through Eq. 3, the average $k$ value obtained for each complex was used. This approach has been shown to give Mn–H distances with an error of less than 4%.[34] A $^1$H NMR spectrum of each sample was recorded at each temperature to ensure that the spectrum was static, and to confirm that no $T_1$ averaging was occurring.

**2.4. Computational Details/Experimental Charge Density Analyses.** DFT calculations were performed with the GAUSSIAN 98/03 program suite[35] using the BPW91 density functional,[36] along with the implemented 6-311G(d,p) basis set.[37] All geometry optimizations (except [HMn(CO)$_5$] **6**: $C_4$ symmetry) were carried out without imposing any symmetry constraints. The reported structures were confirmed as true minima on the respective potential energy surface by calculating analytical frequencies. The computation of Wiberg[38a] and overlap-weighted natural atomic orbital (NAO) bond orders[38b-c] was performed using the algorithm in version 3.1 of the NBO program,[38d] as implemented in GAUSSIAN 03; the topology of $\rho(\mathbf{r})$ was analyzed using the AIMPAC software package.[39] Delocalization indices were calculated using the AIMDELOC01 script developed by C. F. Matta[40] and an approximation suggested by J. Poater *et al.*[41,42] Kohn-Sham orbitals were plotted and analyzed using the program VMD[43] and a Mathematica routine written by M. Presnitz.[44] A modified version of the AIMPAC[39,45] and the XD[46] code has been used to evaluate $S(\mathbf{r},\Omega)$ at the given reference points $\mathbf{r}$ and to perform topological analyses of theoretical and experimental charge density distributions, $\rho(\mathbf{r})$.

Experimental $\rho(\mathbf{r})$ distributions of **2** were based on multipolar refinements published in ref. 19 (see Supporting Material for detailed information).

## 3. Results and Discussion

**3.1. Structure Determination of 1 by Neutron Diffraction.** The molecular structure of **1** as determined by single-crystal neutron diffraction is shown in Figure 1, and salient structural parameters are listed in Table 2 in comparison with those of its fluoro congener **2**, the only other silane $\sigma$-bond complex which has been studied by neutron diffraction.[11] The structural parameters of **1** and **2** are remarkably similar, and both complexes show a close correspondence with other hydrosilane complexes of the [($\eta^5$-C$_5$R$_5$)Mn(CO)L] fragment that have been studied by single-crystal X-ray diffraction.[7] However, complex **1** is the only example containing both a coordinated and an uncoordinated Si–H moiety in the vicinity of the metal atom. Hence, **1** is an ideal benchmark system for analysis of the bonding and structural changes attendant on silane coordination to a transition metal center. Indeed, **1** displays all established features associated with $\sigma$-bond complexation of the Mn center: the coordinated Si–H bond occupies a single site in a three-legged piano stool complex, and the Si–H distance of 1.806(14) Å is about 20% longer than the $r$(Si–H) value in four-coordinate SiH$_4$ [*ca.* 1.4798(4) Å; high resolution IR spectroscopy in the gas phase].[47a] Furthermore, the uncoordinated Si–H bond 1.501(13) Å at the five-coordinate Si center in **1** is slightly lengthened relative to SiH$_4$, but still shorter compared with the Si-H bonds in the five-coordinate [H$_2$SiPh$_3$]$^-$ anion [1.593(2) and 1.602(2) Å; single-crystal X-ray diffraction].[47b]

==============

**Figure 1** near here

==============

The relative orientation of the hydrosilane and the [Cp'Mn(CO)$_2$] fragment is determined by the interplanar angle, $\alpha_{ip}$, spanned by the H(1)–Si–Mn and C(1)–Mn,C(2) moieties (Table 2). Since both the interplanar angles $\alpha_{ip}$ and the (O)C–Mn–C(O) angles are close to 90° in **1** and **2**, we can define a coordinate system at the central manganese atom in straightforward way: the *x*-axis bisects the carbonyl groups, the *y*-axis lies in the plane of the carbonyl ligands and is orthogonal relative to the *x*-vector, while the *z*-axis points toward the midpoint of the coordinating Si–H moiety (Figure 3d). Hence, both carbonyl ligands are directed toward the Mn($d_{xy}$) orbital while the coordinated Si–H moiety is in the optimal orientation favored for back-donation from the metal center, with optimal overlap between the acceptor orbital $\sigma$*(Si–H) and the metal-based $d_{yz}$ orbital.[18] A detailed molecular orbital analysis for **2** is presented in Section 3.3.

==============

**Table 2** near here

==============

It is noteworthy that the uncoordinated Si–H(2) bond also lies in the plane defined by the coordinated Si–H(1) moiety and the central Mn atom [$\tau${H(1)–Mn–Si–H(2)} = 177.3(9)°]. Hence, the geometry at the Si center can appropriately be described as a distorted trigonal bipyramid (TBP); the major distortion being the rather tight H–Si–H angle of 148.5(8)° wrought by the Mn–H interaction. It is instructive, then, to compare the geometry at the Si atom in **1** with that found in two related systems which have been characterized by neutron diffraction. These are the complex [Cp$_2$NbH(SiMe$_2$Cl)$_2$] **4**,[48] which exhibits an inter-ligand hypervalent interaction (IHI) between the hydride and silane moieties bound to Nb, and the hypercoordinate silane anion in the complex [K(18-crown-6)][H$_2$SiPh$_3$] **5**,[47b,49] as depicted in Figure 2.

===============

**Figure 2** near here

===============

Each of these three systems can be considered to contain a HSiL$_4$ moiety in which Si is five-coordinate, and the degree to which this departs from a regular TBP geometry reflects the nature of the bonding at the Si center. Thus, the strong covalent side-on coordination of the Si–H moiety to the Mn center in **1** results in its elongation by 0.31 Å relative to its uncoordinated twin, along with a H–Si–H angle which deviates by more than 30° from linearity. For **4**, the weak interaction between Si and H is reflected in an even longer Si⋯H distance of 2.076(3) Å; again, the Cl–Si–H angle clearly deviates from linearity (by 21.6°). In contrast, the rather electrostatic end-on interaction of one Si–H unit with the alkali metal cation in **5** does not cause significant distortions of the almost regular TBP geometry at silicon.

**3.2. $T_1$ NMR studies of 1 and 2.** To gain an insight into the solution state structure of complexes **1** and **2**, we carried out an extensive series of NMR relaxation studies. This approach has also been applied to other TM hydrides containing metal nuclei with large magnetic moments, such as Re and Nb. Earlier studies of monohydride,[33,34] and of silyl-hydride complexes[48a] have proven the utility of this approach. Applying Eq. 3 to **1** we deduce a $r$(Mn–H) value of 1.56(3) Å, which is in excellent agreement with that obtained by neutron diffraction [1.575(14)Å]. $T_{1min}$ for this complex was measured as 185 ms (at 220 K); this value is characteristic of a classical Mn(I) hydride.[33] Such a conclusion is not unexpected, as the Mn–H distances observed by neutron diffraction for **1** and **2** are comparable to those measured for classical Mn(I) hydrides such as [HMn(CO)$_5$][50] **6** (Table 3). We obtained similar results for **2** [$r_{NMR}$(Mn–H) = 1.56(3) and $r_{ND}$(Mn–H) = 1.569(4) Å; ND = neutron diffraction]; $T_{1min}$ for **2** is slightly higher, at 192 ms, but the overall result is in line with that observed for **1** and the solid-state diffraction data.

==============

**Table 3** near here

==============

Using Eq. 2 to evaluate the Mn–H distances through the $T_{1min}$ data alone, we find values of 1.54(1) and 1.55(1) Å for **1** and **2**, respectively. This excellent agreement indicates that relaxation via MHDDI dominates in these systems. Similar results were found for the Nb(III) monohydride [Cp$_2$NbH(SiMe$_2$Cl)$_2$][48a] (Table 3), in accord with the fact that all of these systems contain hydrides that are fairly distant from any other ligand protons. We conclude that this NMR relaxation technique provides a straightforward and reliable method for investigating the structure of manganese hydrides and silane $\sigma$-bond complexes in solution.

**3.3. MO and charge density analyses.** In order to gain a deeper insight into the electronic structure and the bonding situation in the Mn-H-Si unit in **1**, density functional theory (DFT) calculations were carried out in an attempt to examine the Kohn Sham orbitals of **1**. We were particularly interested to draw comparisons with the related complex [Cp'Mn(CO)$_2$($\eta^2$-HSiCl$_3$)] **3** (characterized by X-ray diffraction),[51] on the basis of electronic structure calculations.[18] Earlier topological analyses of its charge density[52] and photoelectron (PE) spectrum[16a,17] have classified **3** as a nearly complete oxidative silane addition product, with only a negligible residual Si–H interaction while **1** and **2** were identified as silane $\sigma$-bond complexes at an early stage of Si–H bond addition.[16b]

Lichtenberger's description of complex **3** as an oxidative addition product is not borne out by our findings. The salient Kohn Sham orbitals accounting for the Mn($\eta^2$–SiH) bonding in **1** are indeed strikingly similar to those obtained for **3** or the model complex [CpMn(CO)$_2$($\eta^2$-HSiCl$_3$] **3a** (Figure 3) in Lichtenberger's study.[18] Whereas the HOMO of **1** and **3** is an orbital of mainly $d_{xz}$ character (see Supporting Information), the main interactions between Mn and Si, as well as Mn and H, manifest themselves in the lower-lying orbital HOMO-2 of **1** and **3**. Additionally but to a lesser extent also HOMO-6 and HOMO-10 of **1** and **3** (Figure 3), respectively, contribute to the Mn–H bonding. As described by Lichtenberger for **3a**, these MOs can be regarded as arising from interaction between the frontier orbitals of the [CpMn(CO)$_2$] fragment with the corresponding orbitals of the incoming silane ligand.[18,53a] Therefore, HOMO-2 in **1** and **3** incorporates the HOMO of the [CpMn(CO)$_2$] fragment, with its high $d_{yz}$ character on the metal center permitting optimal $\pi$ back-bonding to the ligand, whereas HOMO-6/HOMO-10 in **1** and **3**, respectively, contains the LUMO of the [CpMn(CO)$_2$] fragment, with predominant $d_{z^2}$ orbital character, which behaves as a strong acceptor for the approaching ligand (Figure 3).[53b] Such a simplified description emphasizes the Mn–Si and Mn–H bonding interactions at the expense of any residual Si–H attraction in these systems, as pointed out by Lichtenberger in the case of **3a**.[18,54] However, the remarkable similarity of the Mn($\eta^2$-SiH) bonding orbitals of **1** and **3** argues against classification of **3** as a nearly complete oxidative silane addition product, and that of **1** as a silane $\sigma$-bond complex at an early stage of Si–H bond addition.[16b] The similarity in the density contours of the salient Mn($\eta^2$-SiH) bonding molecular orbitals should be naturally reflected in the charge density picture. Indeed, complexes **1**-**3** display strikingly similar charge density distributions in the Mn($\eta^2$-SiH) moiety upon analysis by the 'Atoms in Molecules' (AIM) approach.[55] Since the electron density $\rho(\mathbf{r})$ at a bond critical point (BCP) provides a quantitative and sensitive measure of the bond strength, we can conclude already at this stage that the Si–H bonding characteristics of **1**-**3** are virtually identical: $(\rho(\mathbf{r}))_{\text{Si-H}}$ = 0.52, 0.53, 0.54 eÅ$^{-3}$ for **1**, **2** and **3**, respectively; Figure 4.

Even the Laplacian of the charge density, $\nabla^2\rho(\mathbf{r})$, a highly sensitive measure of subtle changes in the electronic structures (Figure 4) displays rather similar topologies in the Mn–Si–H bonding region. These results, which are based on a physical observable, are pleasingly supported by our recent experimental charge density study of **2** ($(\rho(\mathbf{r}))_{\text{Si-H}}$ = 0.53(4) eÅ$^{-3}$),[19] and disagree with Lichtenberger's conclusion that the Si–H bond within the Mn($\eta^2$-SiH) moiety is broken in the case of **3** but intact in the case of **1** and **2**.[16] Hence, a unifying bonding description is needed which affords an understanding of the fundamental nature of the factors that control Si-H bond activation in silane σ-bond complexes, both for Schubert-type systems and perhaps also in complexes of early TMs.

=============

**Figure 3** near here

=============

=============

**Figure 4** near here

=============

**3.4. Bonding characteristics of the Mn-Si-H moiety.** The similarities in the electronic structures of **1**-**3** are also reflected by their geometrical parameters. As demonstrated earlier,[19] the superposition of the geometries of the DFT-optimized Mn($\eta^2$-SiH$X$) moieties ($X$ = H, F or Cl; respectively) for **1**-**3**, clearly reveals the close structural relationship between the three complexes, with almost identical Si–H and Mn–H bond distances (Table 4). Only the Mn–Si distance – the third parameter characterizing the [Mn–Si–H] moiety – permits discrimination between **1**, **2** and complex **3**. Hence, addition of the polar Si–H moiety to the Mn center occurs in an *asymmetric* manner, proceeding further along the M–H reaction coordinate, so that bond formation between Mn and the more electronegative H atom in **1**, **2** and **3** reaches an advanced stage whereas that between the metal and the more electropositive Si atom lags behind. Complexes **1**-**3** can each then be classified as products of such an asymmetric oxidative addition, albeit at various stages of Si–Mn bond formation.

To allow for a more quantitative comparison between the bonding in **1**, **2** and **3**, two different kinds of bond order indices are listed in Table 4, along with the corresponding bond distances; *viz.* the Wiberg bond index,[38a] as well as the overlap-weighted natural atomic orbital (NAO) bond order,[38b-c] part of the NBO analysis,[38e] and the so-called delocalization index $\delta$,[40a] which represents the number of electron pairs delocalized between two atoms or – more specifically – between atomic basins in the framework of Bader's AIM description.[55] The values in Table 4 reveal that in both descriptions, there is a small but significant increase in the Mn–Si bond order with an increasing number of electronegative substituents at the silicon center. This is also qualitatively reflected by the lack of a discernable Mn–Si bond path for **1**, in contrast to **2**, **3**[19] (Figure 4) and **3a**[52]. However, for the latter complex the Mn–Si bond path was shown to be rather unstable: Bader *et al.* reported in a theoretical study of model **3a** that it disappears when the Mn–Si separation is increased by only 0.05 Å.[52] Our earlier combined experimental and theoretical charge density study confirms this theoretical result and shows that the Mn–Si and Si–H BCPs and ring critical point (RCP) are proximal in **2** and **3**, being located in a region with a rather flat electron density profile (Figure 4).[19] In particular, the curvatures of the Mn–Si bond path in **2** and the Si–H bond path in **3** lead the RCP and BCP almost to merge into a singularity in $\rho(\mathbf{r})$, a confluence characteristic of a scenario close to bond fission. In contrast, the Mn–H bond displays a pronounced charge density at the BCP, together with an almost linear bond path, which indicates a stable bond (Figure 4). The bond indices of the Mn–H moiety are therefore characteristic of a metal hydride such as [HMn(CO)$_5$] **6** (B.O. bond indices: 0.46/0.29; $\delta$ = 0.67). In contrast, the bridging Si–H bonds are considerably weakened relative to the uncoordinated Si–H bond in **1** (B.O.: 0.88/0.75; $\delta$ = 0.53), but still display significant interaction between both atoms for systems **1**-**3**. These findings are consistent with the earlier inference drawn by Schubert[7] from an analysis of structural changes in a range of silane $\sigma$-bond complexes, and also with recent MO interpretations by Choi *et al*.[56] and Nikonov:[57] In effect, the strength of the Mn($\eta^2$-SiH) interaction is dictated by the degree of Mn–Si bonding, which in turn is strongly influenced by the substituents attached to the silicon center.

However, we demonstrate below that the extent of asymmetric oxidative addition of Si–H to the TM center is particularly influenced by the ligand lying *trans* to the $\eta^2$-SiH moiety, and that this is important for a complete understanding of the bonding in these TM silane complexes.

==============

**Table 4** near here

==============

**3.5. The *trans* influence in Mn($\eta^2$-SiH) bonding.** To avoid convolution of steric and electronic effects in the Mn($\eta^2$-SiH) moiety, we introduce here the new model systems [Cp'Mn(CO)$_2$($\eta^2$-HSiMe$_2$X)] (where X = H **1b**, X = F **2b** and X = Cl **3b**). Geometry optimizations of **2b** and **3b** with the electronegative substituents X (X = Cl, F) in *cis* or *trans* locations, respectively to the Si–H moiety, reveal the *trans* position to be energetically favored (by 1.9 and 2.5 kcal/mol after zero-point correction, respectively; Figure 5 and Supporting Material). In addition, the Si–X bonds in **2b** and **3b** are elongated ($\Delta$(Si-X) = 0.015 and 0.032 Å; respectively) in the *trans* position relative to the Si–X bonds in their *cis* congeners. Such an elongation was first interpreted by Crabtree and Hamilton[58] for the fluoro derivative **2** in terms of $\sigma$(Mn–H) → $\sigma^*$(Si–X) donation, which is more pronounced for X = F or Cl than for X = H ($\Delta$(Si–H) = 0.008 Å in **1b**), since the more electronegative the substituent at Si, the greater the Si character in the corresponding $\sigma^*$(Si–X) orbital, and the greater the degree of $\sigma$(Mn–H) → $\sigma^*$(Si–X) charge transfer (Scheme 2). We note that this bonding description is virtually identical with that subsequently proposed by Nikonov for IHIs in hydrido silyl complexes of niobium and tantalum.[48a,59] Crabtree's bonding model also accords with our previous topological analysis of the experimental and theoretical charge density distributions at the Si–F BCP of **2**.[19] According to this study the $\sigma$(Mn–H) → $\sigma^*$(Si–X) charge transfer weakens the Si–F bond, as indicated by a reduced charge density at the Si–F

BCP (Si–F in SiF$_4$: $\rho(\mathbf{r}_c)$ = 0.92 eÅ$^{-3}$; Si–F in **2**: $\rho(\mathbf{r}_c)$ = 0.74 [0.86(5)] eÅ$^{-3}$; experimental values are in square brackets). Such a *trans* interaction results in a bonding scenario closely similar to the negative (or anionic) hyperconjugation commonly invoked for hypercoordinate silicon compounds. As pointed out by Reed *et al.*, such anomeric effects operate at silicon in spite of its electropositivity, and are responsible for the increased conformational flexibility of silicon compounds relative to their carbon analogues.[60] Thus, it appears to be the predominant electron-withdrawing character of the substituent *trans* to the Si–Mn bond that controls the extent of Mn–Si bonding in Mn($\eta^2$-SiH) complexes. The Mn–Si bond in our benchmark systems **2b** and **3b** is slightly but discernibly shorter (by 0.016 and 0.013 Å; respectively) in the *trans vs.* the *cis* variants, and is also shorter than its counterparts in *trans*-**1b** (0.004 Å), which lacks an electronegative substituent in the controlling *trans* position.

==============

**Scheme 2** near here

==============

==============

**Figure 5** near here

==============

Following this argumentation, such $\sigma$(Mn–H) → $\sigma^*$(Si–X) donation would also be expected to result in an elongation of the Mn–H bond. However, the Mn–H bond experiences only a very slight elongation (0.006 and 0.004 Å in **2b** and **3b**; respectively), when the electron-withdrawing ligand is oriented *trans* to the $\sigma$(Si–H) bond. Nor do the experimental or calculated Mn–H bond distances lend support to bond elongation in **2** relative to **1**. Thus, when X = F; Mn–H = 1.569(4), [1.566] Å, but when X = H; Mn–H = 1.575(14), [1.565] Å (computed values in square brackets). Furthermore, the elongation of the Si–H bond and the shortening of the Mn–Si bond in the *trans* as compared to the *cis* variants cannot be explained on the basis of this simple frontier MO approach, which neglects contributions from other orbitals at both the Mn and Si centers that are involved in the overall interaction. A fuller understanding of the complex and interrelated bonding effects at play in the Mn($\eta^2$–SiH) moiety, which are controlled largely by the electronic influence of an auxiliary ligand *trans* to the activated Si–H bond, requires a more sophisticated bonding model.

**3.6. An improved Mn($\eta^2$–SiH) bonding model.** Inspection of the frontier orbitals of **1** and **3** reveals that the LUMO of the silane moiety in each case displays weak Si–H and Si–$X$ ($X$ = H in **1** and $X$ = Cl in **3**) antibonding character (Figure 6). Interaction of these ligand frontier orbitals with the HOMO of the [Cp'Mn(CO)$_2$] metal fragments produces the characteristic Mn($\eta^2$-SiH) bonding MO (HOMO-2) for complexes **1** and **3** (Figure 3 b, e). Using the established synergic bonding model for this type of $\sigma$-bond interaction,[1] HOMO-2 characterizes the $\pi$ back-donation from the filled $d_{yz}$ metal orbital into a three-center ligand orbital displaying both Si–$X$ and Si–H antibonding character. This results in a simultaneous activation of both the $\eta^2$-coordinating Si–H bond and the Si–$X$ bond in *trans* position. Accordingly, the explanation of the control exerted by the *trans*-oriented $X$ ligand on the geometry of the Mn($\eta^2$–SiH) moiety becomes clear: the higher the electron-withdrawing character of $X$, the greater the Si–$X$ and Si–H bond activation, owing to the increased Mn $\rightarrow$ ligand $\pi$ back-donation. This again allows the hyper-coordinated silicon atom to approach closer to the metal center, in accordance with the geometrical trends displayed by complexes **1**-**3** and model systems **1b**-**3b**. This bonding model might also explain the geometrical features observed in hydrido silyl complexes of early *TM* complexes like [Cp$_2$NbH(SiMe$_2$Cl)$_2$] **4**. Furthermore, in contrast to the IHI model suggested by Nikonov,[48a] this approach does not need to invoke any hypervalent character for the silicon atom, since this MO approach does not necessitate recourse to Pauling's *dsp*$^3$ hybridization model.[48b] Indeed, Mulliken and natural population analyses (NPA)[38b-c] of the DFT wavefunction of complexes **1**-**3** clearly show that the 3*d* silicon orbitals act as polarization functions, but play almost no role in the chemical bonding of the Mn($\eta^2$–SiH) moiety (see Supporting Information). In the next Section, we will outline further support for this delocalized bonding approach by analyzing the source function based on experimental and theoretical data.

==============

**Figure 6** near here

==============

**3.7. Delocalized Mn($\eta^2$-SiH) bonding in the source function picture.** As demonstrated by Bader and Gatti,[61a] it is possible to view the electron density, $\rho(\mathbf{r})$, at any point $\mathbf{r}$ within a molecule or solid to consist of contributions from a local source $LS(\mathbf{r},\mathbf{r}')$ operating at all other points $\mathbf{r}'$ of the space. The local source $LS(\mathbf{r},\mathbf{r}')$ contribution at position vector $\mathbf{r}$ from $\mathbf{r}'$ is expressed as:

$$LS(\mathbf{r},\mathbf{r}') = -\frac{1}{4\pi}\frac{\nabla^2\rho(\mathbf{r}')}{|\mathbf{r}-\mathbf{r}'|},$$

where the Laplacian of the electron density at $\mathbf{r}'$, $\nabla^2\rho(\mathbf{r}')$, acts as a source for the electron density at $\mathbf{r}$ with an efficiency given by the Green's function $|\mathbf{r}-\mathbf{r}'|^{-1}$.[61a] By integrating $LS(\mathbf{r},\mathbf{r}')$ over the regions of space bound by the zero flux surfaces defining the atomic basins $\Omega$,[55] the density may be equated to a sum of atomic contributions, $S(\mathbf{r},\Omega)$:

$$\rho(\mathbf{r}) = \int LS(\mathbf{r},\mathbf{r}')d\mathbf{r}' = \int_\Omega LS(\mathbf{r},\mathbf{r}')d\mathbf{r}' + \sum_{\Omega'\neq\Omega} LS(\mathbf{r},\mathbf{r}')d\mathbf{r}'.$$

The integrated form of the source function (SF) is thus defined as

$$S(\mathbf{r},\Omega) \equiv \int_\Omega LS(\mathbf{r},\mathbf{r}')d\mathbf{r}'$$

and provides a measure of the relative importance of each atom's contribution to the density at a specific reference point $\mathbf{r}$ – a decomposition which allows one to view the properties of the density from a new perspective and which reveals the SF as a powerful tool, able to provide chemical insight in complex bonding scenarios.[61b,g] For example, analysis of the SF has proved very useful in the characterization of hydrogen bonds[61b,c,f] and of weak intra- and intermolecular interactions,[61d] and has also been used to analyze metal-metal interactions in dimeric cobalt complexes[61h] and bimetallic carbonyl complexes[42a] and the interaction of TMs with delocalized $\pi$-hydrocarbyl ligands.[61e]

We have calculated the SF contribution using the BCPs of the Mn($\eta^2$-HSiFPh$_2$) moiety of **2** as reference points **r** (Figure 7). Taking the Mn–H BCP as reference point, the major SF contributions arise from the Mn atom (27% [29%]); experimental values in square brackets) and the H atom (39% [39%]). The contribution of Si is small (3% [3%]), indicating a strongly localized Mn–H bond. Using the Mn–Si BCP as reference, the situation is quite different. Here Mn, Si and H each contribute to a very similar extent to the density at the BCP (13% [16%], 19% [21%], 18% [15%] for Mn, Si and H, respectively). This implies a strongly delocalized bond, and therefore supports the bonding model introduced in the previous Section. The formation of the Mn–Si bond directly affects the Si–H bond due to the $\pi$ back-donation from the metal $d_{yz}$ into the antibonding $\sigma^*$ orbital. For the same reason, we can anticipate a significant contribution from the Mn atom to the charge density at the Si–H BCP. For complex **2**, we find that 7% [6%] of the density at the critical point originates from the metal atom. In line with the above discussion, this contribution increases from **1** (5%) through **2** (6%) to **3** (9%), confirming that the back-donation from Mn to Si increases as the Si center becomes more electropositive. Hence, the SF picture is in pleasing accord with the MO model, and confirms that the Si–Mn bonding is the result of a complex interplay of contributions from all four atoms of the Mn($\eta^2$-HSiF) moiety.

## 4. Conclusions

In contrast to earlier findings, the Mn-Si-H bonding in the complexes [Cp'Mn(CO)$_2$($\eta^2$-HSiHPh$_2$)] **1** and the fluoro derivative [Cp'Mn(CO)$_2$($\eta^2$-HSiFPh$_2$)] **2** is not fundamentally different from that in systems like [Cp'Mn(CO)$_2$($\eta^2$-HSiCl$_3$)] **3**, which possess more than one electronegative ligand at Si. Complex **3** displays a shorter Mn–Si bond with a slightly higher value of $\rho(\mathbf{r})$ at the BCP relative to **2**, but the Si–H or Mn–H interaction is not significantly weaker than that in **1** or **2**. The Mn–Si–H bonding in all of these systems is naturally accommodated by an *asymmetric* oxidatitive addition reaction coordinate in which the Mn–H bond is formed at an early stage, while the establishment of the Mn–Si bond is controlled and enforced by the extent of Mn → $\sigma^*$(*X*-Si–H) $\pi$ back-donation, which increases with increasing electron-withdrawing character of the *X* substituent in the *trans* position to the metal-coordinated Si–H bond. With an increase in this charge transfer, both the $\eta^2$-coordinated Si–H bond and the *trans* Si–*X* bond become activated, while the asymmetry in the Mn–H and Mn–Si bonding is reduced. The salient geometrical and electronic characteristics of the Mn($\eta^2$-*X*SiH) moiety are dictated by the electron-withdrawing character of the *X* substituent *trans* to the coordinated Si–H bond at the hypercoordinate silicon atom, which controls the extent of the $\pi$ back-donation in a synergic bonding situation. The Si-based acceptor orbital in this model is Si–H and Si–*X* antibonding in nature. As the electronegativity of *X* is increased, the energy of this acceptor orbital is lowered and the degree of Mn → $\sigma^*$(*X*-Si-H) back-donation is increased, thereby lengthening both the Si–H and the Si–*X* bonds. This MO interpretation is supported by combined experimental and theoretical charge density analyses, since the source function (SF) unequivocally shows that the Mn–Si bonding arises from a complex interplay of SF contributions from all four atoms of the Mn($\eta^2$-HSiF) moiety. As pointed out by Reed *et al.*, such anomeric effects operate at silicon in spite of its electropositivity, and are responsible for the increased conformational flexibility of silicon compounds relative to their carbon analogues.[60] This might explain at least in part why the corresponding Schubert-type metal alkyl complexes have not been isolated so far. Electron delocalization effects have also been identified as the underlying driving force in the related class of agostic early transition metal alkyls displaying activated C–H bonds,[62] and in lithium organic species characterized by distorted alkyl moieties and short Li···H contacts.[63] Electron delocalization thus appears to be a common driving force behind the structural distortions in these related but distinct types of C–H and Si–H metal complexes.

**Acknowledgment.** We thank Prof. Niimura for providing beam time and his support for the neutron diffraction experiment of **2**. EPSRC (UK), NSERC and CFI (Canada), DFG Germany (SPP1178) and NanoCat (an International Graduate Program within the Elitenetzwerk Bayern) are all gratefully acknowledged for financial support.

**Supporting Information Available:** Cartesian coordinates, Mulliken and NPA analyses, geometrical and topological parameters and atomic charges for **1-3, 1b-3b** and **6**, and detailed $S(\mathbf{r},\Omega)$ analyses of **1-3**. Plots of the HOMO and LUMO for **1-3**; crystallographic data for **1** in CIF format. Details of the multipolar refinements of **2**. This material is available free of charge via the Internet at http://pubs.acs.org.

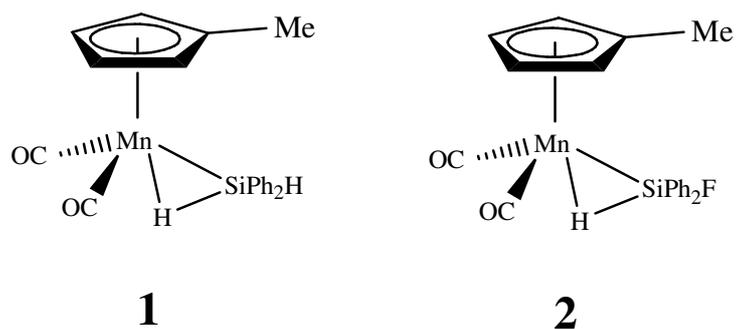

**Scheme 1**

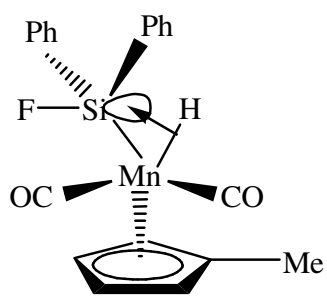

**Scheme 2**

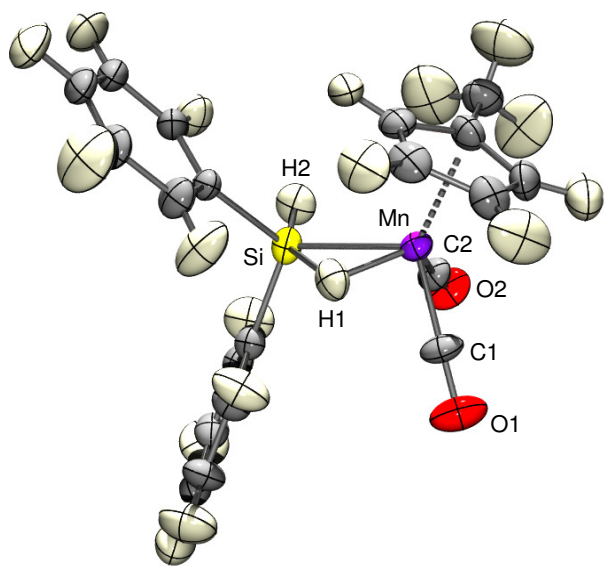

**Figure 1.** Structural model of [Cp'Mn(CO)$_2$($\eta^2$-HSiHPh$_2$)] **1** as determined by single-crystal neutron diffraction. Ellipsoids are drawn at the 30% probability level.

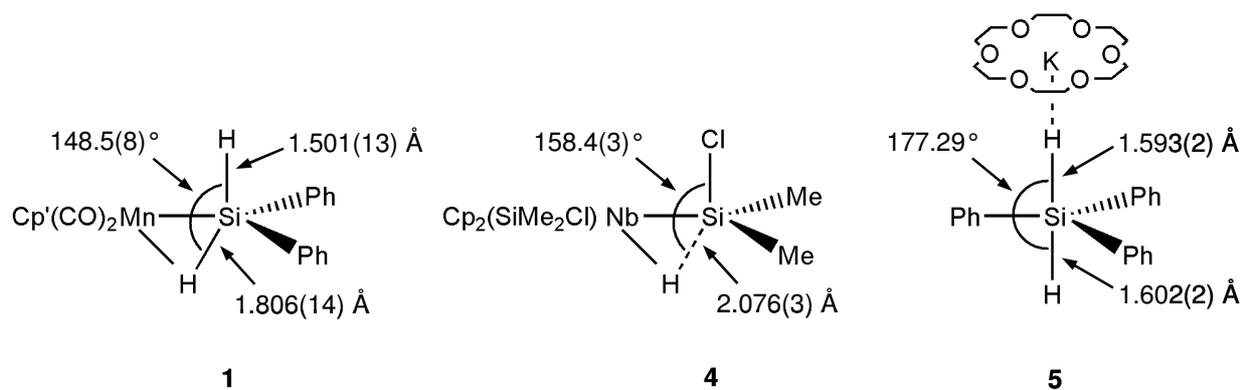

**Figure 2.** Comparison of structural features at the Si center of [Cp'Mn(CO)$_2$($\eta^2$-HSiHPh$_2$)] **1**, [Cp$_2$NbH(SiMe$_2$Cl)$_2$][48a] **4** and [K(18-crown-6)][H$_2$SiPh$_3$][49] **5**, as determined by single-crystal neutron diffraction.

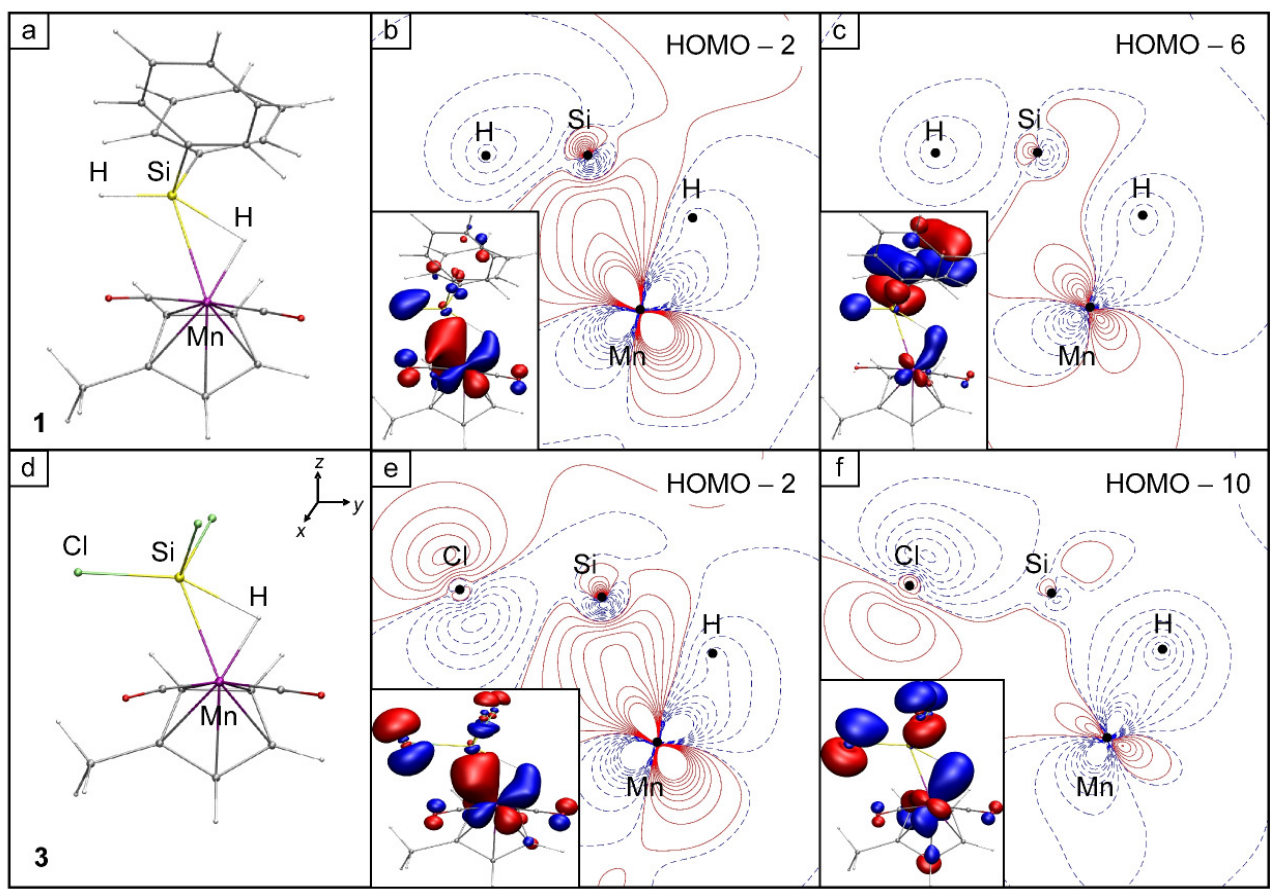

**Figure 3.** Molecular orbitals of [Cp'Mn(CO)$_2$($\eta^2$-HSiHPh$_2$)] **1** (b-c) and [Cp'Mn(CO)$_2$($\eta^2$-HSiCl$_3$)] **3** (e-f) at the BPW91/6-311G(d,p) level of theory. For the orientation of **1** and **3** the Mn($\eta^2$-H-Si) moiety was located in the molecular *y,z* plane with the *z* axis originating at the metal center and pointing toward the midpoint of the Si-H vector. Density contour levels are drawn at $\pm n \times 0.025$ a.u., where $n = \{1,\ldots,8\}$; extra level at 0.0025 a.u.; all iso-surface densities (insets) are displayed at a contour value of $\pm 0.05$ a.u.

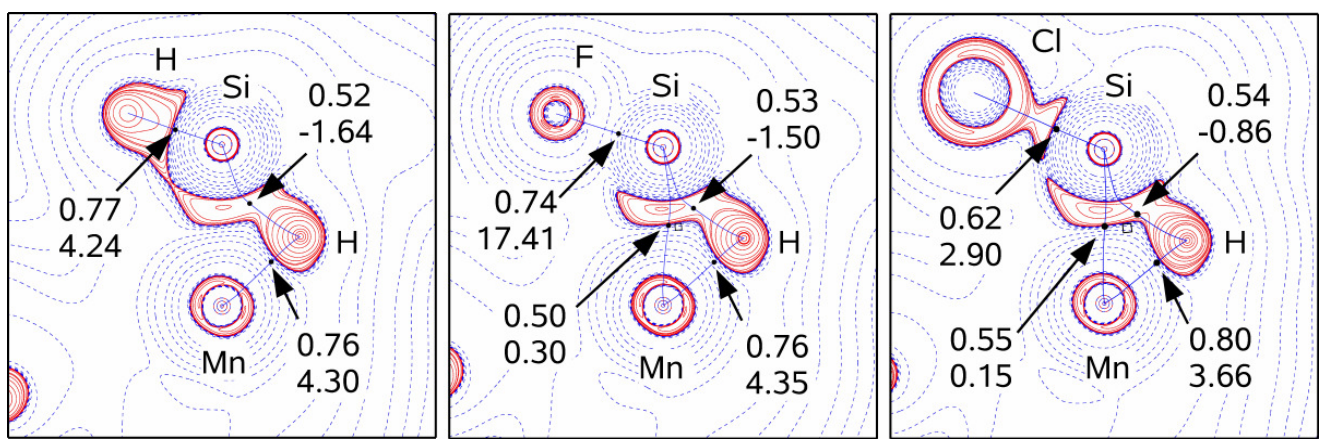

**Figure 4.** $\nabla^2\rho(\mathbf{r})$ contour maps of the electron density of [Cp'Mn(CO)$_2$($\eta^2$-HSiHPh$_2$)] **1**, [Cp'Mn(CO)$_2$($\eta^2$-HSiFPh$_2$)] **2**, and [Cp'Mn(CO)$_2$($\eta^2$-HSiCl$_3$)] **3** in the Mn–H–Si plane. Contour levels are drawn at 0.001, $\pm 2.0 \times 10^n$, $\pm 4.0 \times 10^n$, $\pm 8.0 \times 10^n$ eÅ$^{-5}$, where $n = 0, 3, \pm 2, \pm 1$; extra levels at 2.4, 2.8, 15, 150, 180, 700 eÅ$^{-5}$; negative and positive values are marked by solid and dashed lines, respectively. BCPs and RCPs are marked by closed circles and squares, respectively; the bond paths are shown by solid lines; $\rho(\mathbf{r})/\nabla^2\rho(\mathbf{r})$ at the critical points in [eÅ$^{-3}$/ eÅ$^{-5}$], respectively.

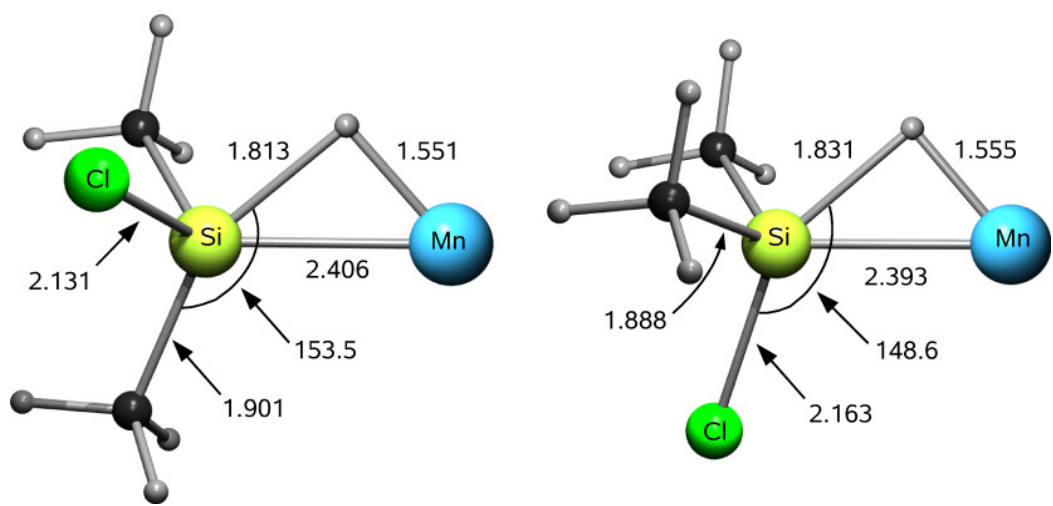

**Figure 5.** BPW91/6-311G(d,p) optimised *cis* and *trans* orientations of the Mn($\eta^2$-SiHClMe$_2$) moiety in [Cp'Mn(CO)$_2$($\eta^2$-HSiMe$_2$Cl)] **3b**; salient bond distances and angles in Å or degrees, respectively.

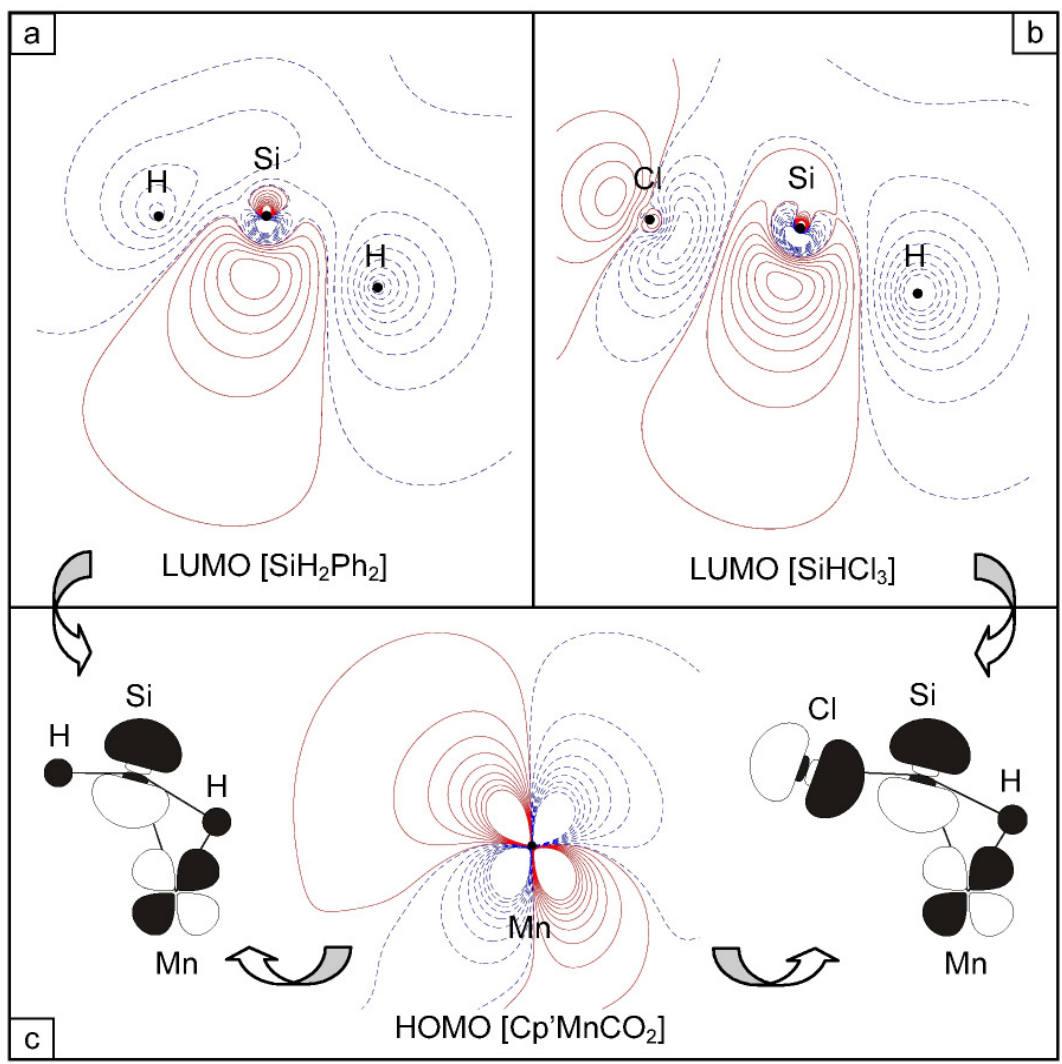

**Figure 6.** (a) Frontier molecular orbitals of the silyl moiety (LUMO) of (a) [Cp'Mn(CO)$_2$($\eta^2$-HSiHPh$_2$)] **1**, (b) [Cp'Mn(CO)$_2$($\eta^2$-HSiCl$_3$)] **3** and (c) the metal fragment Cp'Mn(CO)$_2$ (HOMO) in the molecular *y,z* plane; for a definition of the molecular orientation and the contour values employed, see Figure 3. (c) Schematic drawing of the composition of the HOMO-2 in **1** and **3** by these frontier orbitals. According to our synergic bonding description, HOMO-2 represents the Mn → ligand $\pi$ back-donation.

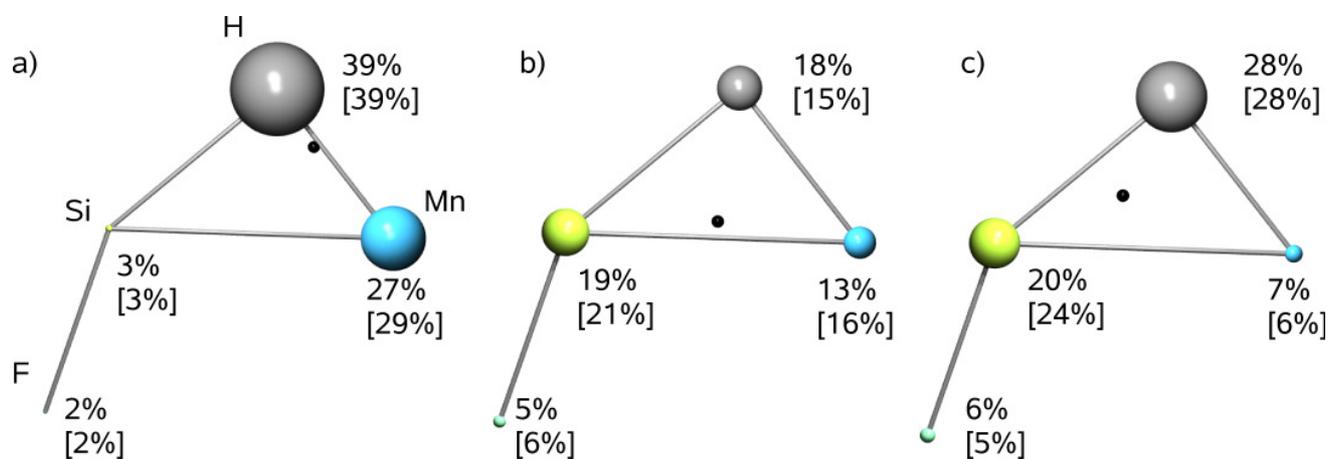

**Figure 7.** Percentage atomic source contributions to the electron density at (a) the Mn–H BCP, (b) the Mn–Si BCP, and (c) the Si–H BCP of [Cp'Mn(CO)$_2$($\eta^2$-HSiFPh$_2$)] **2** determined from the experimental electron density distribution and (in square brackets) as obtained by DFT calculations. The position of the reference points are indicated by black spheres. The volume of the spheres is proportional to the source contributions from the respective atomic basins.

**Table 1.** Crystallographic Data for [Cp'Mn(CO)$_2$($\eta^2$-HSiHPh$_2$)] **1**.

| | |
|---|---|
| chemical formula | C$_{20}$H$_{19}$MnO$_2$Si |
| $M_r$ | 374.38 |
| crystal system | monoclinic |
| color, habit | yellow, plate |
| size (mm) | 3.0 × 2.3 × 2.0 |
| space group | $P2_1/n$ |
| $a$ (Å) | 8.4750(5) |
| $b$ (Å) | 15.1016(10) |
| $c$ (Å) | 14.8686(10) |
| $\beta$ (deg) | 105.425(3) |
| $V$ (Å$^3$) | 1834.4(2) |
| $Z$ | 4 |
| $T$ (K) | 293(1) |
| $\rho_{calc}$ (g·cm$^{-3}$) | 1.356 |
| $\lambda$ (Å) | 1.23 |
| $\mu$ (mm$^{-1}$) | 0.155 |
| $\theta$ range (deg) | 3.39-38.09 |
| data collected ($h, k, l$) | -8;8, -15;15, -14;14 |
| no. of rflns measured | 12727 |
| no. of unique rflns | 1920 |
| no. of observed rflns; $I > 2\sigma(I)$ | 1165 |
| $R_1{}^a$ (obsd), $wR_2{}^b$ (all) | 0.072, 0.176 |
| GooF | 0.949 |
| no. of parameters refined | 388 |

$^a$ $R_1 = \Sigma(|F_o| - |F_c|)/\Sigma|F_o|$. $^b$ $wR_2 = \{\Sigma[w(F_o^2 - F_c^2)^2]/\Sigma[w(F_o^2)^2]\}^{1/2}$.

**Table 2.** Salient bond distances (Å), bond angles (deg) and torsional angles (deg) for [Cp'Mn(CO)$_2$($\eta^2$-HSiHPh$_2$)] **1**[11] and [Cp'Mn(CO)$_2$($\eta^2$-HSiFPh$_2$)] **2** as determined by single-crystal neutron diffraction.

| Parameter | 1 | 2 |
|---|---|---|
| Mn–H(1) | 1.575(14) | 1.569(4) |
| Si–H(1) | 1.806(14) | 1.802(5) |
| Mn–Si | 2.391(12) | 2.352(4) |
| Si–H(2),F | 1.501(13) | 1.634(3) |
| C(1)–Mn–C(2) | 87.7(5) | 89.7(1) |
| H(1)–Mn–Si | 49.1(5) | 50.0(2) |
| Mn–Si–H(1) | 41.2(5) | 41.8(1) |
| Mn–H(1)–Si | 89.7(7) | 88.2(2) |
| H(1)–Si–H(2),F | 148.5(8) | 148.8(2) |
| C(1)–Mn–H(1)–Si | -129.8(6) | -140.9[a] |
| C(2)–Mn–Si–H(1) | 135.5(8) | 126.6[a] |
| H(1)–Mn–Si–H(2),F | 177.3(9) | 174.1[a] |
| $\alpha_{ip}$[b] | 84.2(8) | 83.7[a] |

[a] Errors omitted when not reported in the original paper.
[b] $\alpha_{ip}$ denotes the interplanar angle spanned by the H(1)–Si–Mn and C(1)–Mn–C(2) moieties.

**Table 3.** Comparison of M–H bond lengths (in Å) obtained for a series of hydride and silane σ-bond complexes studied by neutron diffraction, $T_1/T_{1s}$ NMR measurements and DFT calculations.

| Complex | $r$(M–H) NMR | $r$(M–H) ND | $r$(M–H) DFT |
|---|---|---|---|
| [(Cp')Mn(CO)$_2$(η$^2$-HSiHPh$_2$)] **1** | 1.56(3)[d] | 1.575(14)[c] | 1.565[c] |
| [(Cp')Mn(CO)$_2$(η$^2$-HSiFPh$_2$)] **2** | 1.56(3)[a] | 1.569(4)[b] | 1.566[c] |
| [HMn(CO)$_5$] **6** | 1.65(5)[e] | 1.601(16)[f] | 1.571[c] |
| [Cp$_2$NbH(SiMe$_2$Cl)$_2$] **4** (central hydride isomer) | 1.78(1)[g] | 1.816(8)[g] | 1.811[g] |
| [Cp$_2$NbH$_2$(SiMe$_2$Cl)] (central silyl isomer) | 1.71(1)[g] |  | 1.739[g] |
| [Cp$_2$NbH$_2$(SiMe$_2$Cl)] (lateral silyl isomer) | 1.68(1)[g] 1.74(1)[g] |  | 1.745[g] 1.793[g] |

[a] This work; obtained using a $T_{1min}$ value of 191.55 ms and an averaged $k$ value of 0.2. [b] Ref. 11. [c] This work. [d] This work; obtained using a $T_{1min}$ value of 185.49 and an averaged $k$ value of 0.4. [e] Ref. 33. [f] Ref. 50. [g] Ref. 48a.

**Table 4.** Comparison of bond distances (in Å; theoretical values in square brackets) and bond order indices for the Mn-H-Si moiety in the complexes [Cp'Mn(CO)$_2$($\eta^2$-HSiXPh$_2$)] (**1**: X = H; **2**: X = F), [Cp'Mn(CO)$_2$($\eta^2$-HSiCl$_3$)] **3** and [Cp'Mn(CO)$_2$($\eta^2$-HSiClMe$_2$)] **3b**; B.O.: Wiberg/overlap-weighted NAO bond order; $\delta$: delocalization index.

| Complex | Mn–Si distance | B.O. | $\delta$ | Mn–H distance | B.O. | $\delta$ | Si–H$_b$ distance | B.O. | $\delta$ |
|---|---|---|---|---|---|---|---|---|---|
| **1** | 2.391(12)$^a$ | | | 1.575(14)$^a$ | | | 1.806(14)$^a$ | | |
| | [2.417] | 0.43/0.31 | 0.42 | [1.565] | 0.41/0.29 | 0.64 | [1.804] | 0.36/0.45 | 0.28 |
| **2** | 2.352(4)$^b$ | | | 1.569(4)$^b$ | | | 1.802(5)$^b$ | | |
| | [2.367] | 0.48/0.35 | 0.47 | [1.566] | 0.41/0.30 | 0.63 | [1.817] | 0.34/0.44 | 0.28 |
| **3** | 2.254(1)$^c$ | | | –$^d$ | | | –$^d$ | | |
| | [2.310] | 0.56/0.39 | 0.60 | [1.555] | 0.44/0.31 | 0.63 | [1.841] | 0.30/0.40 | 0.28 |
| *cis*-**3b**$^e$ | [2.406] | 0.48/0.34 | | [1.551] | 0.43/0.31 | | [1.813] | 0.34/0.43 | |
| *trans*-**3b** | [2.393] | 0.50/0.35 | | [1.555] | 0.43/0.31 | | [1.831] | 0.32/0.41 | |

$^a$ This work. $^b$ Ref. 11. $^c$ X-ray data; ref. 51. $^d$ No neutron data available. $^e$ For a definition of the *cis* and *trans* orientations of the Mn($\eta^2$-SiHClMe$_2$) moiety see Figure 5.

SYNOPSIS TOC (Word Style "SN_Synopsis_TOC"). If you are submitting your paper to a journal that requires a synopsis graphic and/or synopsis paragraph, see the Guide, Notes, Notice, or Instructions for Authors that appear in each publication's first issue of the year and the journal's homepage for a description of what needs to be provided and for the size requirements of the artwork.

[1] Kubas G. J. *Metal Dihydrogen and σ-Bond Complexes*; Kluwer Academic/Plenum Publishers: New York, **2001**.

[2] Kubas, G. J. *J. Organomet. Chem.* **2001**, *635*, 37.

[3] McGrady, G. S.; Guilera, G. *Chem. Soc. Rev*. **2003**, *32*, 383.

[4] Crabtree, R. H. *Angew. Chem., Int. Ed. Engl.* **1993**, *32*, 789.

[5] Hoyano, J. K.; Elder, M.; Graham W. A. G. *J. Am. Chem. Soc.* **1969**, *91*, 4568.

[6] See, for example: (a) Geftakis, S.; Ball, G. E. *J. Am. Chem. Soc.* **1998**, *120,* 9953. (b) Castro-Rodriguez I.; Nakai, H.; Gantzel P.; Zakharov, L. N.; Rheingold A. L.; Meyer, K. *J. Am. Chem. Soc*. **2003**, *125*, 15734.

[7] Schubert, U. *Adv. Organometal. Chem.* **1990**, *30*, 151.

[8] Corey, J. Y.; Braddock-Wilking, J. *Chem. Rev*. **1999**, *99,* 175.

[9] Lin, Z. *Chem. Soc. Rev*. **2002**, *31*, 239.

[10] Lachaize, S.; Sabo-Etienne, S. *Eur. J. Inorg. Chem*. **2006**, 2115.

[11] Schubert, U.; Ackermann, K.; Wörle, B. *J. Am. Chem. Soc*. **1982**, *104*, 7378.

[12] Atheaux, I. ; Delpech, F. ; Donnadieu, B. ; Sabo-Etienne, S. ; Chaudret, B. ; Hussein, K. ; Barthelat, J. C. ; Braun, T. ; Duckett, S. B. ; Perutz, R. N. *Organometallics* **2002**, *21*, 5347.

[13] Nikonov, G. I. *J. Organomet. Chem.* **2001**, *635*, 24.

[14] Bakhmutov, V. I.; Vorontsov, E. V. *Rev. Inorg. Chem.* **1998**, 183.

[15] Gusev, D. G.; Kuhlman, R. L.; Renkema, K. B.; Eisenstein, O.; Caulton, K. G. *Inorg. Chem.* **1996**, *35*, 6775.

[16] (a) Lichtenberger, D. L.; Rai-Chaudhuri, A. *J. Am. Chem. Soc.* **1989**, *111*, 3583. (b) Lichtenberger, D. L.; Rai-Chaudhuri, A. *J. Am. Chem. Soc.* **1990**, *112*, 2492.

[17] The actual system studied by photoelectron spectroscopy was [CpMn(CO)$_2$($\eta^2$-HSiCl$_3$)] (**3a**), with an ($\eta^5$-C$_5$H$_5$) ligand instead of ($\eta^5$-C$_5$H$_4$Me), as in **3**. The differences in Mn($\eta^2$-SiH) bonding between these complexes should be small, so we have chosen to include complex **3** in our study, since it has been characterized by X-ray diffraction (ref. 51).

[18] Lichtenberger, D. L. *Organometallics* **2003**, *22*, 1599.

[19] Scherer, W.; Eickerling, G.; Tafipolsky, M.; McGrady, G. S.; Sirsch, P.; Chatterton, N. P. *Chem. Commun.* **2006**, 2986.

[20] Erington, R. J. *Advanced Practical Inorganic and Metallorganic Chemistry*, Nelson Thornes: New York, **1997**.

[21] Schubert, U.; Scholz, G.; Müller, J.; Ackermann, K.; Wörle, B.; Stansfield, R. *J. Organomet. Chem.* **1986**, *306*, 303.

[22] Tanaka, I.; Kurihara, K.; Chatake, T.; Niimura, N. *J. Appl. Cryst.* **2002**, *35*, 34.

[23] Niimura, N.; Karasawa, Y.; Tanaka, I.; Miyahara, J.; Takahashi, K.; Saito, H.; Koizumi, S.; Hidaka, M. *Nucl. Instrum. & Methods Phys. Res.* **1994**, *A349*, 521.

[24] Tanaka, I.; Niimura, N.; Mikula, P. *J. Appl. Cryst.* **1999**, *32*, 525.

[25] Otwinowski, Z.; Minor, W. *Processing of X-ray Diffraction Data Collected in Oscillation Mode*, Methods in Enzymology, Volume 276: Macromolecular Crystallography, part A; Carter, C. W., Sweet, R. M., Eds.; Academic Press, **1997**, p. 307-326.

[26] Blessing, R. H. *Acta Cryst.* **1995**, *A51*, 33.

[27] Sheldrick, G. M. SHELXL-97, Program for Crystal Structure Refinement, University of Göttingen, Göttingen, Germany, **1997**.

[28] Sears, V. F. *Neutron News* **1992**, *3*, 26.

[29] Spek, A. L. *J. Appl. Cryst.* **2003**, *36*, 7.

[30] Farrugia, L. J. *J. Appl. Cryst.* **1997**, *30*, 565.

[31] Abragam, A. *The Principles of Nuclear Magnetism*; Oxford University Press: New York, **1971**.

[32] Sugiura, M.; Takao, N.; Fujiwara, H. *Mag. Res. Chem.* **1988**, *26*, 1051.

[33] Farrar, T. C.; Quinting, G. R. *J. Phys. Chem.* **1986**, *90*, 2834.

[34] Gusev, D. G.; Nietlispach, D.; Vymenits, A. B.; Bakhmutov, V. I.; Berke, H. *Inorg. Chem.* **1993**, *32*, 3270.

[35] Frisch, M. J.; Trucks, G. W.; Schlegel, H. B.; Scuseria, G. E.; Robb, M. A.; Cheeseman, J. R.; Montgomery, J. A., Jr.; Vreven, T.; Kudin, K. N.; Burant, J. C.; Millam, J. M.; Iyengar, S. S.; Tomasi, J.; Barone, V.; Mennucci, B.; Cossi, M.; Scalmani, G.; Rega, N.; Petersson, G. A.; Nakatsuji, H.; Hada, M.; Ehara, M.; Toyota, K.; Fukuda, R.; Hasegawa, J.; Ishida, M.; Nakajima, T.; Honda, Y.; Kitao, O.; Nakai, H.; Klene, M.; Li, X.; Knox, J. E.; Hratchian, H. P.; Cross, J. B.; Adamo, C.; Jaramillo, J.; Gomperts, R.; Stratmann, R. E.; Yazyev, O.; Austin, A. J.; Cammi, R.; Pomelli, C.; Ochterski, J. W.; Ayala, P. Y.; Morokuma, K.; Voth, G. A.; Salvador, P.; Dannenberg, J. J.; Zakrzewski, V. G.; Dapprich, S.; Daniels, A. D.; Strain, M. C.; Farkas, O.; Malick, D. K.; Rabuck, A. D.; Raghavachari, K.; Foresman, J. B.; Ortiz, J. V.; Cui, Q.; Baboul, A. G.; Clifford, S.; Cioslowski, J.; Stefanov, B. B.; Liu, G.; Liashenko, A.; Piskorz, P.; Komaromi, I.; Martin, R. L.; Fox, D. J.; Keith, T.; Al-Laham, M. A.; Peng, C. Y.; Nanayakkara, A.; Challacombe, M.; Gill, P. M. W.; Johnson, B.; Chen, W.; Wong, M. W.; Gonzalez, C.; Pople, J. A. *Gaussian 03*, Revision B.05, Gaussian, Inc., Pittsburgh PA, **2003**.

[36] (a) Becke, A. D. *Phys. Rev.* **1988**, *A38*, 3098. (b) Perdew, J. P.; Wang, Y. *Phys. Rev.* **1992**, *B45*, 13244.

[37] (a) McLean, A. D.; Chandler, G. S. *J. Chem. Phys.* **1980**, *72*, 5639. (b) Krishnan, R.; Binkley, J. S.; Seeger, R.; Pople, J. A. *J. Chem. Phys.* **1980**, *72*, 650. (c) Wachters, A. J. H. *J. Chem. Phys.* **1970**, *52*, 1033. (d) Hay, P. J. *J. Chem. Phys.* **1977**, *66*, 4377. (e) Raghavachari, K.; Trucks, G. W. *J. Chem. Phys.* **1989**, *91*, 1062.

[38] (a) Wiberg, K. A. *Tetrahedron* **1968**, *24*, 1083. (b) Reed, A. E.; Weinhold, F. *J. Chem. Phys.* **1983**, *78*, 4066. (c) Reed, A. E.; Weinstock, R. B.; Weinhold, F. *J. Chem. Phys.* **1985**, *83*, 735. (d) Glendening, E. D.; Reed, A. E.; Carpenter, J. E.; Weinhold F. *QCPE Bull.* **1990**, *10*, 58. (e) Weinhold, F. "Natural Bond Orbital Methods," in, Encyclopedia of Computational Chemistry, Schleyer, P. v.R.; Allinger, N. L.; Clark, T.; Gasteiger, J.; Kollman, P. A.; Schaefer III, H. F.; Schreiner P. R. (Eds.), John Wiley & Sons, Chichester, UK, **1998**, Vol. 3, pp. 1792.

[39] (a) Biegler-König, F. W.; Bader, R. F. W.; Tang, T. *J. Comput. Chem.* **1982**, *3*, 317. (b) Cheeseman, J. R.; Keith, T. A.; Bader, R. F. W. *AIMPAC program package*, McMaster University, Ontario, Canada, **1997**.

[40] (a) Fradera, X.; Austen, M. A.; Bader, R. F. W. *J. Phys. Chem. A.* **1999**, *103*, 304. (b) Matta, C. F. AIMDELOC01, QCPE0802, QCPE, *Indiana University*, **2001**.

[41] Poater, J.; Solà, M.; Duran, M.; Fradera, X. *Theor. Chem. Acta.* **2002**, *107*, 362. According to Poater *et al.* the delocalization indices $\delta(\Omega,\Omega')$ of DFT wavefunctions can be calculated using an approximate formula that makes use of an HF-like second order exchange density matrix. According to a recent study by Gatti *et al.* (ref 42a) this approximation affords $\delta(\Omega,\Omega')$ values which are very close to the HF ones if the HF and DFT optimized geometries are similar, although it erroneously implies that the electron-pair density matrix can be constructed, within DFT, using the same simple formalism valid for the HF method. For more recent examples employing this concept see also ref 42b.

[42] (a) Gatti, C.; Lasi, D. *Faraday Discuss.* **2007**, *135*, 55; (b) Hebben, N.; Himmel, H.-J.; Eickerling, G.; Herrmann, C.; Reiher, M.; Herz, V.; Presnitz, M.; Scherer, W. *Chem. Eur. J.* **2007**, *13*, 10078.

[43] Humphrey, W.; Dalke, A.; Schulten, K. *J. Molec. Graphics* **1996**, *14*, 33.

[44] (a) Wolfram Research, Inc. *Mathematica Version 5.2;* Wolfram Research, Inc.: Champaign, IL, 2005; (b) Presnitz, M.; Mayer, F.; Herz, V.; Eickerling, G.; Scherer, W. "calc.lap.nb", Universität Augsburg (Lehrstuhl CPM), 2007 Mathematica Script for the Analysis of Charge Density Distributions.

[45] Gatti, C.; Bertini L. *Acta Cryst*. **2004**, *A60*, 438.

[46] XD2006 (version 5.42) - a computer program for multipole refinement, topological analysis of charge densities and evaluation of intermolecular energies from experimental or theoretical structure factors. Volkov, A.; Macchi, P.; Farrugia, L. J.; Gatti, C.; Mallinson, P.; Richter, T.; Koritsanszky, T.; **2006**.

[47] (a) Boyd, D. R. *J. Chem. Phys.* **1955**, *23*, 922; (b) Bearpark, M. J.; McGrady, G. S.; Prince, P. D.; Steed, J. W. *J. Am. Chem. Soc.* **2001**, *123*, 7736.

[48] (a) Bakhmutov, V. I.; Howard, J. A. K.; Keen, D. A.; Kuzmina, L. G.; Leech, M. A.; Nikonov, G. I.; Vorontsov, E. V.; Wilson, C. C. *J. Chem. Soc. Dalton Trans*. **2000**, 1631. (b) Burdett, J. "Chemical Bonds: A Dialog," in, Inorganic Chemistry a Text Book Series", Meyer, G.; Nakamura A.; Woolins, D. (Eds.), John Wiley & Sons, Chichester, UK, **1998**, pp. 47.

[49] McGrady, G. S.; Prince, P. D.; Steed, J. W.; Gutmann, M. J., unpublished results.

[50] La Placa, S. J.; Hamilton, W. C.; Ibers, J. A.; Davison, A. *Inorg. Chem.* **1969**, *8*, 1928.

[51] Schubert, U.; Ackermann, K.; Kraft, G.; Wörle, B. *Z. Naturforsch., B.: Chem. Sci.* **1983**, *38*, 1488.

[52] Bader, R. F. W.; Matta, C. F.; Cortés-Guzmán, F. *Organometallics* **2004**, *23*, 6253.

[53] (a) Schilling, B. E. R.; Hoffmann, R.; Lichtenberger, D. L. *J. Am. Chem. Soc.* **1979**, *101*, 585. (b) A more detailed inspection of the HOMO-2 and HOMO-6/HOMO-10 contours in **1** and **3**, respectively, suggests that the metal-centered atomic orbitals can be better described in terms of linear combinations

of the Mn($d_{yz}$) and Mn($d_{z^2}$) orbitals: (HOMO-2: $d_{yz}$ + $d_{z^2}$) and (HOMO-6/HOMO-10: $d_{z^2}$-$d_{yz}$). Significant mixing of these metal centered *d* orbitals was also found by Lichtenberger for **3a** (ref 18) by analyzing localized orbitals.

[54] As shown by Weinhold, Curtiss, Reed et al. ((a) Foster, J. P.; Weinhold, F. *J. Am. Chem. Soc.* **1980**, *102*, 7211. (b) Reed, A. E.; Weinhold, F. *J. Chem. Phys.* **1983**, *78*, 4066. (c) Reed, A. E.; Curtiss, L. A.; Weinhold, F. *Chem. Rev.* **1988**, *88*, 899), molecular orbitals can be transformed into a set of localized few-center natural bond orbitals (NBOs), which resemble the Lewis-like bonding pattern of localized electron pairs. The localized NBOs of **3** based on our own calculations are strikingly similar to the localized orbitals computed by Lichtenberger for **3a** in Ref 18. We note, however, that the occupancy of the two Mn−Si and Mn−H NBOs deviates significantly from the ideal value of two electrons (1.78 and 1.69, respectively), and that the latter NBO displays a significant delocalization toward the silicon atom. Accordingly, the description of the Mn($\eta^2$−Si-H) bonding in terms of these two localized NBOs alone is misleading and insufficient.

[55] Bader, R. F. W. *Atoms in Molecules: A Quantum Theory*; Clarendon Press: Oxford, **1994**.

[56] Choi, S-H.; Feng, J.; Lin, Z. *Organometallics*, **2000**, *19*, 2051.

[57] Nikonov, G. I. *Adv. Organomet. Chem.* **2005**, *53*, 217.

[58] Crabtree, R. H.; Hamilton, D. G. *Adv. Organomet. Chem.* **1988**, *28*, 299.

[59] (a) Nikonov, G. I.; Kuzmina, L. G.; Vyboishchikov, S. F.; Lemenovskii, D. A.; Howard, J. A. K. *Chem. Eur. J.* **1999**, *5*, 2497; (b) Nikonov, G. I.; Mountford, P.; Ignatov, S. K.; Green, J. C.; Cooke, P. A.; Leech, M. A.; Kuzmina, L. G.; Razuvaev, A. G.; Rees, N. H.; Blake, A. J.; Howard, J. A. K.; Lemenovskii, D. A. *Dalton Trans.* **2001**, 2903.

[60] Reed, A. E.; Schade, C.; Schleyer, P. v. R.; Kamath, P. V.; Chandrasekhar J. *J. Chem. Soc., Chem. Commun.* **1988**, 67.

[61] (a) Bader, R. F. W.; Gatti, C. *Chem. Phys. Lett.* **1998**, *287*, 233. (b) Gatti, C.; Cargnoni, F.; Bertini, L. *J. Comput. Chem.* **2003**, *24*, 422. (c) Overgaard, J.; Schiott, B.; Larsen, F. K.; Iversen, B. B. *Chem. Eur. J.* **2001**, *7*, 3756. (d) Gatti, C.; Bertini L. *Acta Cryst.* **2004**, *A60*, 438; (e) Farrugia L.J., Evans, C. ; Tegel, M. *J. Phys. Chem. A,* **2006**, 110, 7952; (f) Sørensen, J.; Clausen, H. F.; Poulsen, R. D., Overgaard J.; Schiøtt, B. *J. Phys. Chem. A*, **2007**, *111*, 345; (g) Bertini, L.; Cargnoni, F.; Gatti, C. *Theor. Chem. Acc*, **2007**, *117*, 847; (h) Overgaard, J.; Clausen, H.F.; Platts,J. A.; Iversen B.B. *J. Am. Chem. Soc.* **2008**, *130*, 3834.

[62] (a) Haaland, A.; Scherer, W.; Ruud, K.; McGrady, G. S.; Downs, A. J.; Swang, O. *J. Am. Chem. Soc.* **1998**, *120,* 3762; (b) Scherer, W.; Priermeier, T.; Haaland, A.; Volden, H. V.; McGrady, G. S.; Downs, A. J.; Boese, R.; Bläser, D. *Organometallics*, **1998**, *17*, 4406; (c) Scherer, W.; Sirsch, P.; Shorokhov, D.; Tafipolsky, M.; McGrady, G. S. Gullo, E. *Chem. Eur. J*. **2003**, *9*, 6057; (d) Scherer, W.; McGrady, G.S. *Angew. Chem. Int. Ed.* **2004**, *43*, 1782.

[63] (a) Scherer, W. ; Sirsch, P. ; Grosche, M. ; Spiegler, M. ; Mason, S. A. ; Gardiner, M. G. *Chem. Commun.* **2001**, 2072; (b) Scherer, W.; Sirsch, P.; Shorokhov, D.; McGrady, G. S.; Mason, S. A.; Gardiner, M. G. *Chem. Eur. J.* **2002**, **8**, 2324.